\documentstyle[psfig]{mn}
\newcommand{\bgr}{\bibitem[\protect\citename{dummy }1893]{dum}}
\newcommand{\etalc}{et~al.}
\begin{document}
\title[XBLs]
{Variations in the broadband spectra of BL Lac objects: millimetre 
observations of an X-ray-selected sample}

\author[J. A. Stevens and W. K. Gear ]
{J.~A.~Stevens\footnotemark\addtocounter{footnote}{-1} and W. K. Gear\footnotemark
\\
Mullard Space Science Laboratory, University College London, Holmbury
St. Mary, Dorking, Surrey, RH5 6NT  
}
\date{draft 1.0}

\maketitle
\begin{abstract}
Observations at millimetre wavelengths are presented for a
representative sample of 22 X-ray-selected BL Lac objects (XBLs). This sample
comprises 19 High-energy cutoff BL Lacs objects (HBLs), 1 Low-energy cutoff BL
Lac object (LBL) and 2 `intermediate' sources. Data for LBLs, which are mostly
radio selected BL Lac objects (RBLs) are taken from the
literature. It is shown that the radio--millimetre spectral indices 
of HBLs ($\bar\alpha_{5-230}=-0.29\pm0.15$) are slightly steeper than those of
the LBLs ($\bar\alpha_{5-230}=-0.19\pm0.14$). A correlation exists between
$\alpha_{5-230}$ and 230~GHz luminosity. While this correlation could be an 
artefact of comparing two populations of BL Lac objects with intrinsically
different radio properties it is also consistent with the predictions of
existing unified schemes which relate BL Lac objects to Fanaroff-Riley class I
radio galaxies.

The HBLs have significantly flatter submillimetre--X-ray spectral indices 
($\bar\alpha_{230-X}=-0.68\pm0.07$) than the LBLs
($\bar\alpha_{230-X}=-1.08\pm0.06$) although the two intermediate sources also 
have intermediate values of $\alpha_{230-X}\sim-0.9$.  
It is argued that this difference cannot
be explained entirely by the viewing angle hypothesis and requires a difference
in physical source parameters.
The $\alpha_{230-X}$ values for the HBLs are close to the canonical value
found for large samples of radio sources and thus suggest that
synchrotron radiation is the mechanism that produces the X-ray emission. 
As suggested by Padovani \& Giommi, 
the inverse-Compton mechanism is likely to dominate in the LBLs
requiring the synchrotron spectra of these sources to steepen or cutoff at
lower frequencies than those of the HBLs.

\end{abstract}

\begin{keywords}
galaxies: active --
BL Lacertae objects: general --
radiation mechanisms: non-thermal --
radio continuum: galaxies 
\end{keywords}
%
\footnotetext{E-mail: jas@mssl.ucl.ac.uk(JAS); wkpg@mssl.ucl.ac.uk(WKG)}

\section{Introduction}

More BL Lac objects have now been discovered with X-ray surveys than
with radio surveys. By definition, objects selected by the two
techniques share many common properties such as compact core-jet
structures, lack of or weak emission lines, rapid and large amplitude
optical and radio variability, high and variable linear polarization
and flat radio spectra. These general properties are universally
attributed to the presence of a relativistic jet that is angled
towards the line-of-sight (e.g. Blandford \& Rees 1978).

There are, however, several differences between XBLs and RBLs. Most
strikingly, XBLs have significantly lower apparent radio luminosities than RBLs
(e.g. Stocke et al. 1990) whilst
both classes have similar apparent X-ray luminosities (Sambruna et
al. 1994 and references therein). Additionally, XBLs
typically have lower optical polarizations and optical variability
amplitudes (Schwartz et al. 1989;
Jannuzi, Green \& French 1993a; Jannuzi, Smith \& Elston 1993b),
different parsec-scale radio properties (Kollgaard et al. 1996a),
may exhibit negative
evolution (Morris et al. 1991; Perlman et al. 1996a; Bade et al. 1998) 
while RBLs exhibit
positive evolution (Stickel et al. 1991) and have a larger galaxy
fraction in their optical images (Wurtz 1994) and near-infrared colours
(Gear 1993a). Furthermore, XBLs may exhibit steeper X-ray spectra than
RBLs but this result is only strictly true when comparing
high-energy cutoff BL Lac objects (HBLs) which are mostly XBLs with
low-energy cutoff BL Lac objects (LBLs) which are mostly RBLs
(Padovani \& Giommi 1996; Urry et al. 1996; see also Section 2).

Two competing models have been devised that attempt to
explain at least some of these differences: (1) the beaming model
in which XBLs are seen at larger angles to the line-of-sight than are
RBLs (e.g. Ghisellini \& Maraschi 1989; Padovani \& 
Urry 1991; Ghisellini et al. 1993) and (2) a model in which the cutoff
frequency due to synchrotron radiative losses is at higher frequencies for XBLs
(UV--X-ray) than for RBLs (optical--IR; Padovani \& Giommi 1996 and
references therein). 

For the latter model, the difference in radio power between the two classes can
be explained as a selection effect whereas the beaming model accounts for this
difference in a natural way, although the jets have to accelerate between the
X-ray emitting region and the radio emitting region in order to reproduce the
observed behaviour. 

Recent studies suggest that the beaming model in
combination with an intrinsic spread of physical jet properties may provide the
best match to the available data (Sambruna, Maraschi \& Urry 1996;
Georganopoulos \& Marscher 1998; Fossati et al. 1998).

This paper presents new millimetre wavelength observations of a
representative sample of 22 XBLs. A previous study that included data
in this band (Gear 1993b) hinted that XBLs and RBLs exhibit similar
radio--millimetre spectral indices but different millimetre--X-ray
spectral indices, indicating that any differences between the two
classes will be more apparent in the X-ray band. These results were
based on a very small number of sources, however, and so the aim of
this study is to provide a similar analysis for a larger number of
objects. Throughout this paper we assume the Hubble constant,
H$_0=50$ Kms$^{-1}$Mpc$^{-1}$, $\Omega_0=0$ and spectral indices are written as
$S_{\nu}\propto\nu^{\alpha}$.

\section{Sample}

BL Lac objects occupy a distinct region of the
$\alpha_{ox}-\alpha_{ro}$ plane (Stocke et al. 1990). 
In fact, it was initially found that the RBLs
and XBLs occupied two distinct regions but as more XBLs were discovered 
the separation became less pronounced (see below). The distribution is shown in
Fig.~1 which includes RBLs from the `1 Jansky' sample (Stickel et al. 1991) and
XBLs predominantly from the {\em Einstein} Medium Sensitivity Survey (EMSS;
e.g. Perlman et al. 1996a), the {\em Einstein} Slew Survey (Perlman et
al. 1996b) and the {\em ROSAT} all-sky survey (Kock et al. 1996; Nass et al. 1996;
Bade et al 1998). Also included are sources from the Deep X-ray Radio Blazar
Survey (DXRBS) which were found by correlating the {\em ROSAT}
WGACAT database with several radio surveys (Perlman et al. 1998). Some BL Lac
objects discovered in the {\em ROSAT} surveys tend to populate the 
$\alpha_{ox}-\alpha_{ro}$
plane in the region between the `1 Jansky' RBLs and the {\em
Einstein} XBLs and can be considered intermediate objects. Note also that 
many of the DXRBS sources have colours typical of the RBLs.

We selected predominantly those sources with 4.85 GHz fluxes greater than 
50~mJy but included fainter sources in an attempt to populate the
$\alpha_{ox}-\alpha_{ro}$ plane in a uniform manner. 
Those sources with detections at millimetre wavelengths are shown as filled
symbols in Fig.~1; millimetre XBL observations are either
from this study (see Table~1) or from Gear (1993b) whilst the millimetre 
RBL observations are from Gear et al. (1994). 

\begin{table*}
\centering                                                                     
\def\baselinestretch{1}                                     
\caption[dum]{\small Summary of the observations}
\vspace*{0.25in}
\small
\begin{tabular}{lccccrr}\hline
source & UT date & 2 mm  & 1.35 mm & 0.85 mm &$\alpha_{5-230}$ &
$\alpha_{230-X}$ \\
       & &(mJy) & (mJy)   & (mJy) & &\\ 
(1)& (2) & (3) & (4) & (5) & (6) & (7)\\ \hline                              
RXJ 00079+4711   & 19970921 & $3\sigma<17.3$ & \ldots & \ldots & \ldots & \ldots \\ 
                & 19970624 & \ldots & 3$\sigma<16.3$ & \ldots & $<-0.34$ & $>-0.84$\\
                & 19980208 & \ldots & $3\sigma<7.5$ & \ldots & $<-0.56$ & $>-0.78$ \\
                & 19980531 & \ldots & $3\sigma<7.5$ & \ldots & $<-0.56$ & $>-0.78$ \\
1ES 0033+595     & 19970921 & 34.3$\pm$8.1 & 25.0$\pm$4.9 & \ldots& $-0.24$ & $-0.64$ \\
4C 47.08        & 19980207 & \ldots & 328.6$\pm$46.8 & \ldots & $-0.50$ &
$-1.03$\\
                & 19980531 & 553.6$\pm$57.2 & 395.9$\pm$43.5 & \ldots & $-0.45$
& $-1.04$ \\
1ES 0323+022     & 19980309 & \ldots & 13.6$\pm$2.9 & \ldots & $-0.29$ & $-0.52$
\\
                & 19980531 & \ldots & 9.5$\pm$2.4 & \ldots & $-0.39$ & $-0.52$ \\
1ES 0414+009     & 19980309 & \ldots & 24.2$\pm$3.6 & \ldots & $-0.28$ & $-0.63$\\
1ES 0446+449     & 19970815 & $3\sigma<12.1$ & \ldots & \ldots & \ldots & \ldots \\ 
                & 19970921 & \ldots & 3$\sigma<14.3$ & \ldots & $<+0.18$ & $>-0.69$\\
1ES 0502+675     & 19980310 & \ldots & 13.1$\pm$3.1 & \ldots & $-0.24$ &  $-0.63$\\
                & 19980311 & \ldots & 18.2$\pm$3.3 & \ldots & $-0.15$ & $-0.65$ \\
EXO 0556.4$-3838$& 19970921 & 27.4$\pm$7.0 & \ldots & \ldots &\ldots & \ldots \\
                & 19980310 & \ldots & 3$\sigma<12.1$ & \ldots & $<-0.45$ & $>-0.48$ \\
1ES 0647+250     & 19970815 & 36.3$\pm$8.5 & \ldots & \ldots & \ldots & \ldots \\
                & 19970921 & \ldots & 33.4$\pm$6.0 & \ldots &$-0.19$ & $-0.65$ \\
1ES 0715$-$259   & 19970921 & $3\sigma<12.0$ & 3$\sigma<6.5$ & \ldots &$<-0.15$ & $>-0.65$\\ 
1ES 0806+524     & 19980310 & & 60.0$\pm$6.8 & \ldots & $-0.28$ & $-0.64$ \\  
1H 0829+089      & 19970921 & 75.7$\pm$9.7 & $3\sigma<8.4$ & \ldots & \ldots & $-0.57$\\
                & 19980310 & \ldots & 14.1$\pm$2.5 & \ldots & $-0.52$ & $-0.60$ \\
                & 19980424 & 43.0$\pm$6.7 & 22.9$\pm$3.8 & \ldots & $-0.40$ & $-0.63$ \\
1ES 1101$-$232     & 19980505 & 16.1$\pm$5.2 & 14.9$\pm$4.2 & \ldots & $-0.39$ & $-0.57$ \\
1ES 1212+078     & 19980505 & 3$\sigma<27.4$ & 17.7$\pm$3.6 & \ldots & $-0.44$ &
$-0.76$ \\
                 & 19980706 & \ldots & 16.7$\pm$3.3 & \ldots & $-0.45$ &
$-0.76$ \\ 
MS 1229.2+6430   & 19980505 & \ldots & 12.6$\pm$4.0 & \ldots & $-0.31$ & $-0.64$ \\
                 & 19980706 & \ldots & $3\sigma<11.8$ & \ldots & $<-0.33$ &
$>-0.64$ \\
RXJ 12302+2512   & 19980505 & 225.4$\pm$27.3 & 208.0$\pm$23.5 & \ldots &
$-0.18$ & $-0.92$ \\
EXO 1415.6+2557  & 19980510 & 3$\sigma<13.1$ & 3$\sigma<12.3$ & \ldots &
$<-0.39$ & $>-0.58$ \\
RXJ 16442+4546   & 19970702 & 17.7$\pm$4.4 & 17.1$\pm$2.9 & \ldots & $-0.49$ & $-0.79$ \\
4U 1722+11       & 19970605 & 118.8$\pm$23.7 & \ldots & \ldots &\ldots & \ldots \\
                & 19980207 & 204.9$\pm$21.3 & 188.2$\pm$19.3 & 182.8$\pm$21.0 &
$+0.18$ & $-0.76$ \\
1H 1730+500      & 19980207 & 57.2$\pm$9.0 & 52.0$\pm$5.7 & \ldots & $-0.29$ & $-0.74$ \\
EXO 1811.7+3143  & 19980207 & 46.4$\pm$8.3 & 40.7$\pm$5.1 & \ldots & $-0.25$ & $-0.94$\\ 
1ES 1959+650     & 19970815 & 144.0$\pm$19.3 & 149.0$\pm$24.8 &\ldots & $-0.14$ & $-0.73$ \\
MS 2143.4+0704   & 19970815 & 14.3$\pm$4.7 & $3\sigma<21.8$ & \ldots &$<-0.22$ & $>-0.79$\\
                & 19980208 & \ldots & 16.7$\pm$3.0 & \ldots & $-0.29$ & $-0.77$ \\
                & 19980531 & \ldots & 11.8$\pm$2.5 & \ldots & $-0.38$ & $-0.74$ \\
                & 19980607 & 14.8$\pm$4.0 & \ldots & \ldots & \ldots & \ldots \\
1ES 2344+514     & 19970624 & 78.9$\pm$11.0 & 54.1$\pm$10.4 & \ldots &$-0.36$ & $-0.74$ \\
                & 19970815 & 79.0$\pm$10.8 & \ldots & \ldots & \ldots & \ldots \\ 
\hline
\end{tabular}
\end{table*}

Two of the sources, 1ES 0446+449 and 1ES 0715$-$4259, have radio
structures typical of radio galaxies (Perlman et al. 1996b) and are not
considered as XBLs in the following analysis. Furthermore, it has been
pointed out by Giommi \& Padovani (1994) and Padovani \& Giommi (1995a) that
the XBL/RBL distinction is not based on physical parameters but only on
the selection band and indeed several BL Lac objects are common to
both samples. These authors proposed that BL Lac objects can be
split into HBLs and LBLs based on the ratio of X-ray to radio 
flux density ($f_x/f_r$). More recent results point to a continuous range of
peak frequencies, making the HBL/LBL distinction itself obsolete 
(e.g. Padovani 1999)
but since this paper is largely concerned with a comparison of BL Lac objects
selected from the two ends of the radio luminosity function 
the HBL/LBL approach is also considered here. The ratio
log$(f_x/f_r)$ (X-ray fluxes measured at 1 KeV and radio fluxes at 5 GHz; 
both quantities in Jy) is calculated for both the `1 Jansky'
sample and for the objects observed in Table~1. The LBL sample is defined as
comprising those objects with log$(f_x/f_r)<-6.25$, the HBL sample has
log$(f_x/f_r)>-5.75$ and an `intermediate' sample, defined by 
$-6.25<$log$(f_x/f_r)<-5.75$ contains only two sources, namely RXJ
12302+2512 and EXO 1811.7+3143.

It turns out that, apart from the two intermediate sources, all of the XBLs 
considered in this paper are HBLs with the exception of 4C~47.08 which is a 
LBL and all of the RBLs are LBLs with the exception of Mrk501 which is a HBL. 
In Section~4 comparison is made between the LBL and HBL samples with comment on
the relation of these to the intermediate sample but because of the large
overlap, the same general results apply to the RBL/XBL samples. 

\begin{figure}
\setlength{\unitlength}{1in}
\begin{picture}(3.5,3.5)
\includegraphics{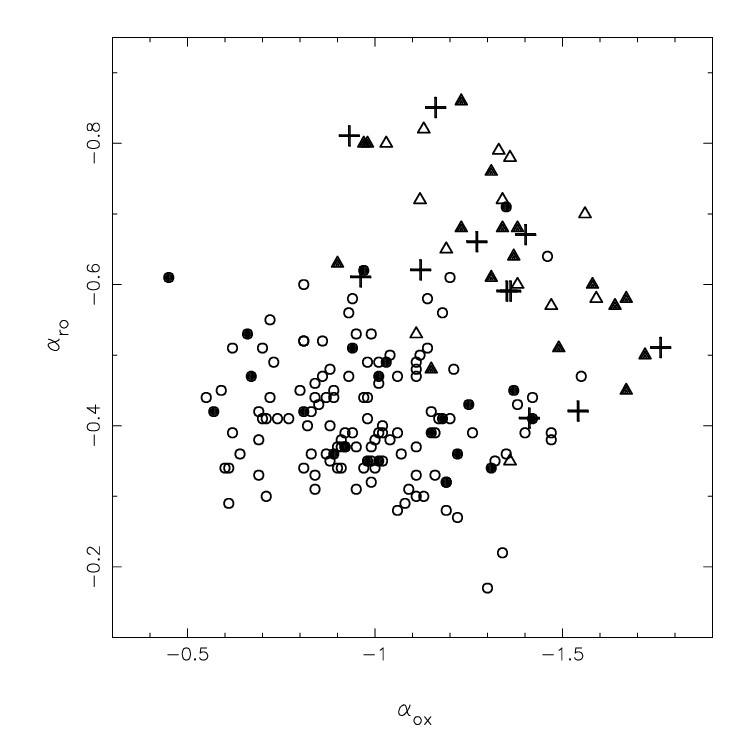}
\end{picture}
\caption[dum]{The distribution of BL Lac objects on the
$\alpha_{ox}-\alpha_{ro}$ plane. XBLs are represented by circles and
`1 Jansky' RBLs are represented by triangles. Those sources highlighted with
filled symbols have detections at 1.35~mm ($\sim$230~GHz). For completeness, 
new BL Lac objects
from the DXRBS survey (see text) are shown as crosses; millimetre observations
were not made for any of these sources.}
\end{figure}

\section{Observations}

\begin{figure*}
\setlength{\unitlength}{1in}
\begin{picture}(7.0,7.0)
\includegraphics{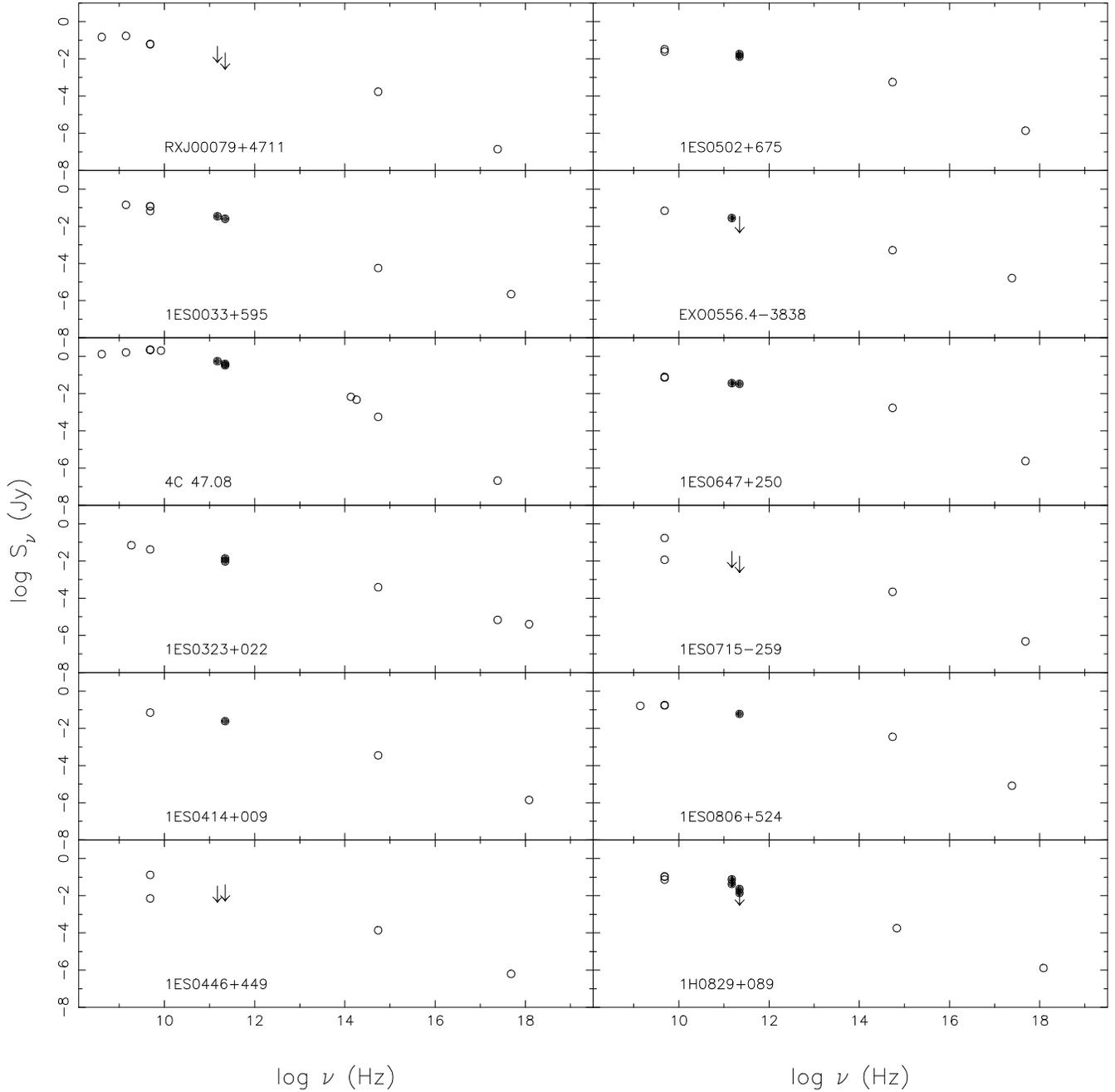}
\end{picture}
\caption[dum]{Broad-band spectra of the BL Lac objects. Millimetre data
presented in this work are shown as filled symbols or upper limits.}
\end{figure*}

\begin{figure*}
\setlength{\unitlength}{1in}
\addtocounter{figure}{-1}
\begin{picture}(7.0,7.0)
\includegraphics{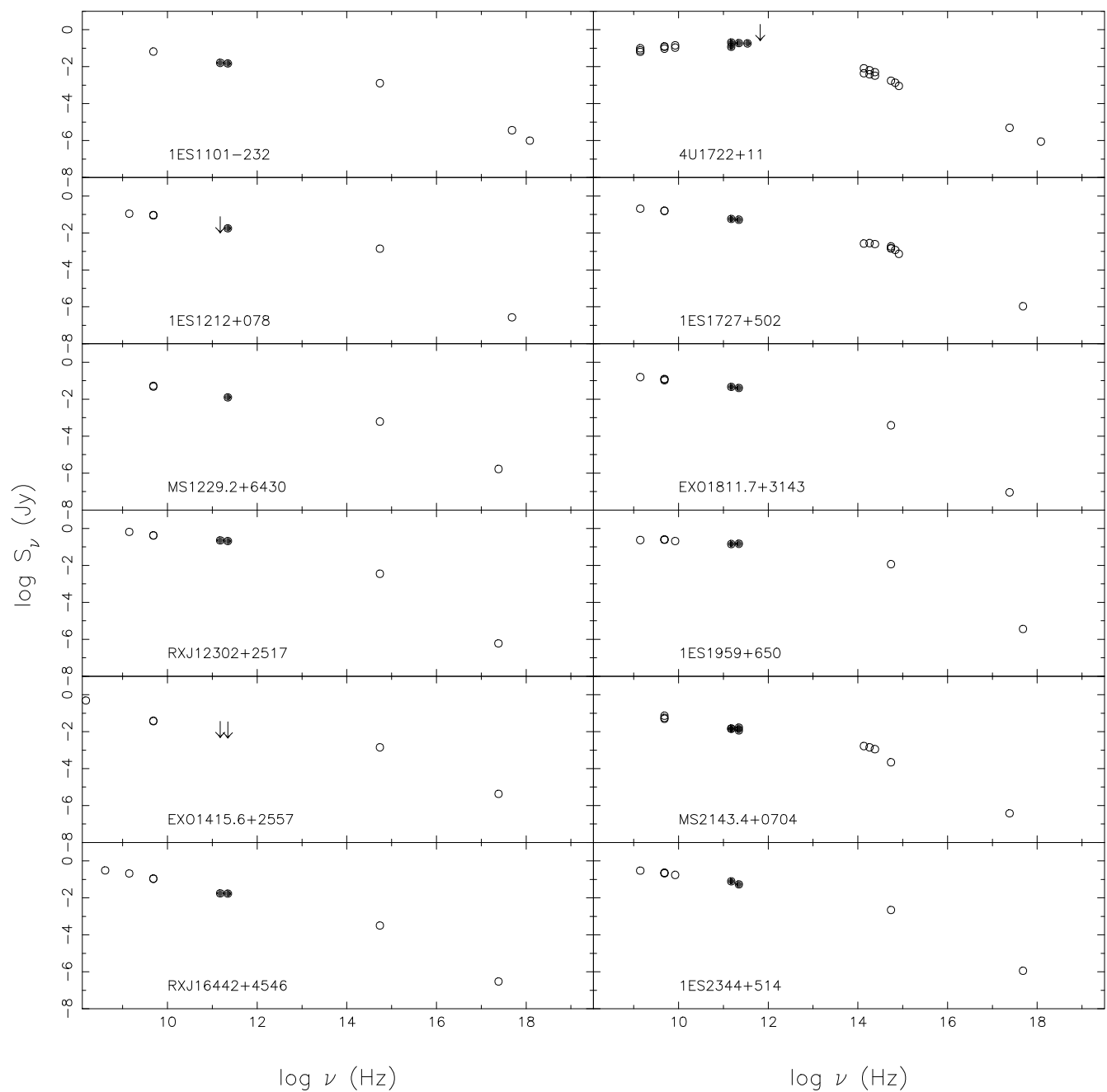}
\end{picture}
\caption[dum]{{\em - continued}}
\end{figure*}

The observations were made with the new submillimetre camera, SCUBA
(Gear et al. 1998, in preparation; Holland et al. 1998,1999), on the 15-m James
Clerk Maxwell Telescope (JCMT), Mauna Kea, Hawaii. SCUBA,
operating at 75~mK, is optimised for mapping observations at
submillimetre wavelengths but also has single bolometers operating at 2.0,
1.35 and 1.1~mm (the 1.1~mm detector was inoperable at the time of this work)
corresponding approximately to 150, 230 and 270 GHz respectively. The NEFDs
at 2.0 and 1.35 mm are typically measured as 120 and 60 mJyHz$^{-1/2}$
respectively (including chopping) and these values are largely 
insensitive to the
atmospheric opacity. Photometric observations can also be made with
the arrays, usually with the central pixel, and in this case the
remaining bolometers can be used to subtract the sky emission which is
correlated across the 2.3 arcmin field of view (Ivison et al. 1998;
Jenness, Lightfoot \& Holland  1998). Under average to poor conditions
the NEFD at 0.85~mm (345~GHz) is $\sim$ 90--200 mJyHz$^{-1/2}$. The
standard SCUBA photometry mode was used for the observations. A small
square map consisting of nine pixels and centred on the source 
is made with an integration time
of one second per pixel, the telescope then nods into the other beam 
where the procedure is repeated. The nine-point maps are
then averaged to produce one photometric observation taking 18 seconds
to complete. The pixel spacing was 3 arcsec at 2.0 and 1.35~mm and 2
arcsec at 0.85~mm and the chop throw was 60 arcsec at $\sim$7.8~Hz.

The observations were typically made under relatively
poor observing conditions, i.e. high levels of precipitable water
vapour (pwv) and atmospheric refraction (sky noise). Thus almost all
data were taken at 2.0 and 1.35~mm to optimise the sensitivity
and reduce the effects of signal-loss due to sky noise and pointing errors
(both giving variations of less than 3 arcsec whilst the beam sizes
are in excess of 20 arcsec at these wavelengths). The atmospheric
opacity was monitored regularly with skydips at both 0.85~mm using
SCUBA and at 1.3 mm with the nearby Caltech Submillimeter Observatory
(CSO) sky monitor. Standard relations were used to convert these opacities to
those relevant to the adopted SCUBA filters. The instrumental gain was
measured on each night with observations of the planets Mars and/or
Uranus. These gains are stable to within 10 per cent at 1.35~mm but may be
elevation dependent at 2.0~mm with a total variation of order 20 per cent. We
have assumed a calibration uncertainty of 10 per cent for all observations and
these are added in quadrature with the uncertainty derived
from the measured signal-to-noise ratio. Data reduction was performed
with the standard SCUBA reduction package and is analogous to that
employed with previous instrumentation.

\section{Results}

The observational results are summarized in Table~1 where column (1)
has the source name, (2) the UT date of the observation, (3), (4) and
(5) the measured fluxes at 2.0, 1.35 and 0.85~mm, (6) the two-point
spectral index, $\alpha_{5-230}$ and (7) the two-point
spectral index, $\alpha_{230-X}$. See below for the definition of these
quantities. Errors on the two-point spectral indices are not quoted since,
because the data are non-simultaneous, it is impossible to quantify
the effect of source variability.

Fig.~2 shows the broad band spectra of those objects listed in
Table~1. These spectra were constructed with non-simultaneous data
taken from the literature but because of the broad frequency range
considered, variability will have only a marginal effect on the
overall spectral shapes. The two radio galaxies mentioned in Section~2
are included in this figure and it should be noted that the radio
spectra include fluxes for both the compact and extended emission, the
latter providing the dominant contribution. Neither source was
detected at millimetre wavelengths. 

In general, the millimetre data fall close to a straight line connecting the
4.85~GHz point with the optical point and the millimetre flux is
typically less than the radio flux as found for LBLs (Gear et
al. 1994) and is thus likely to be optically thin unless the spectra
have a complex structure.

Columns (6) and (7) of Table~1 list the two-point spectral indices between
5~GHz (or more strictly 4.85~GHz) and 230~GHz ($\alpha_{5-230}$) and
between 230~GHz and the X-ray band ($\alpha_{230-X}$). The radio and
X-ray data, which are not contemporaneous with the millimetre data, 
were taken from the literature (Padovani \& Giommi 1995b and reference
therein). These spectral index distributions for the HBL sample are shown in 
the upper panels of Figs.~3 and 4 where two sources detected previously at
230~GHz (Gear 1993b) are also included, namely MS 1402.3+0416 and 
MS 1534.2+0148 (both objects fulfill the HBL criterion defined in Section 2). 

The lower panels of these figures show the
corresponding distributions for the LBL sample. 
The radio and millimetre data used to calculate the LBL
spectral indices (and the HBL Mrk 501) were taken from Gear et al. 
(1994) and are
quasi-simultaneous (taken within $\sim$4 weeks) whilst the X-ray data
(at 1~KeV) are
from Urry et al. (1996) or, for 1514$-$241 and 2200+420 only, from
Worrall \& Wilkes (1990). Fig.~3 includes multiple epochs for
individual sources since source variability can significantly affect the
derived $\alpha_{5-230}$ values even when the two fluxes are measured
quasi-simultaneously (see Gear et al. 1994). In the case of non-detections,
only the deepest upper limit is plotted. However, Fig.~4 only
includes the first millimetre epoch of observation for each source
taken from Gear et al. (1994) or Table~1. In this case, since variability will
have little effect on $\alpha_{230-X}$ due to the large separation in
frequency space, inclusion of multiple epochs would bias the statistics.

\begin{figure}
\setlength{\unitlength}{1in}
\begin{picture}(3.5,2.7)
\includegraphics{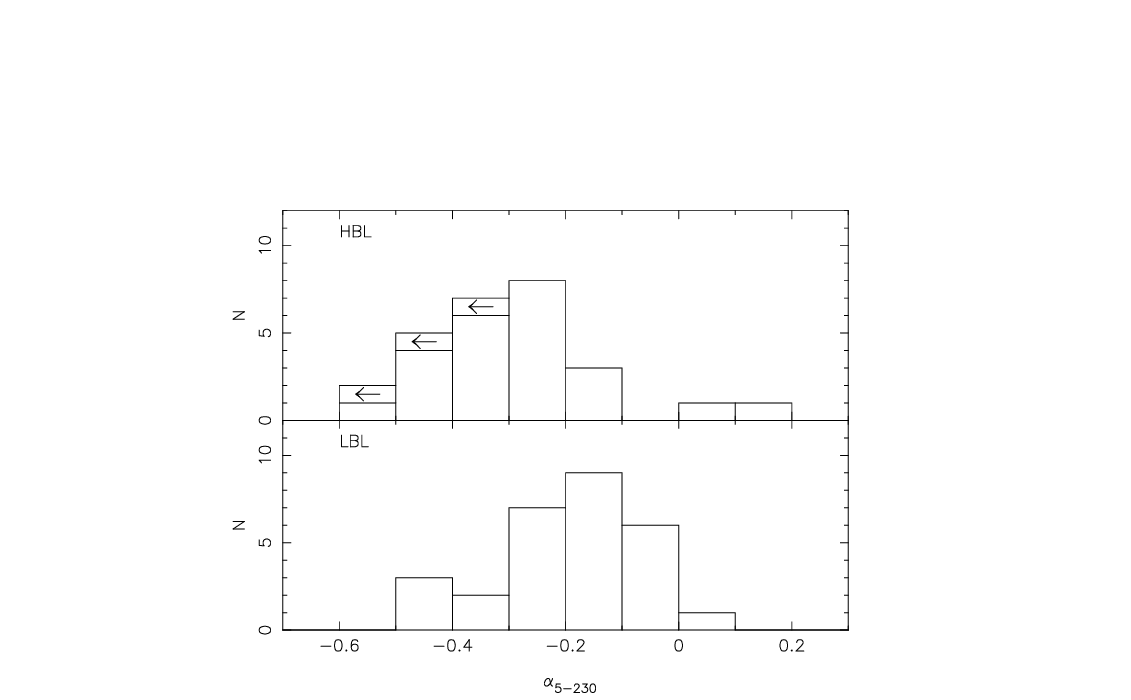}
\end{picture}
\caption[dum]{Histograms showing the 5 to 230 GHz spectral index
($\alpha_{5-230}$) distribution for the HBLs (top) and LBLs (bottom).}
\end{figure}

\begin{figure}
\setlength{\unitlength}{1in}
\begin{picture}(3.5,2.7)
\includegraphics{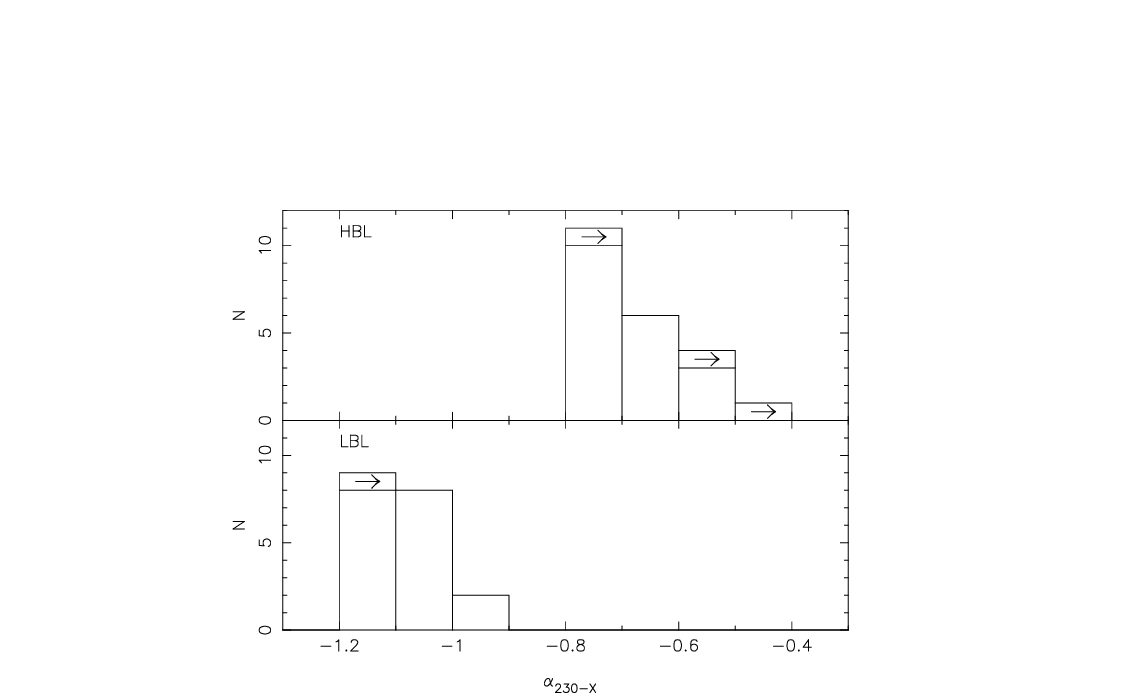}
\end{picture}
\caption[dum]{Histograms showing the 230 GHz to X-ray spectral index
($\alpha_{230-X}$) distribution for the HBLs (top) and LBLs (bottom).}
\end{figure}

The mean values of $\alpha_{5-230}$ are $-0.29\pm0.15$ for the HBLs
and $-0.19\pm0.14$ for the LBLs.
A Kolmogorov-Smirnov (KS) test shows that the $\alpha_{5-230}$
distributions are different at the $99$ per cent confidence level, or if the
limits are assumed to be detections, $99.9$ per cent confidence level,
suggesting that the HBLs have slightly steeper spectra. 
Before concluding that this difference is an intrinsic property it is
first necessary to investigate selection effects; for
example, is $\alpha_{5-230}$ correlated with either redshift or
luminosity?

\begin{figure}
\setlength{\unitlength}{1in}
\begin{picture}(3.5,2.5)
\includegraphics{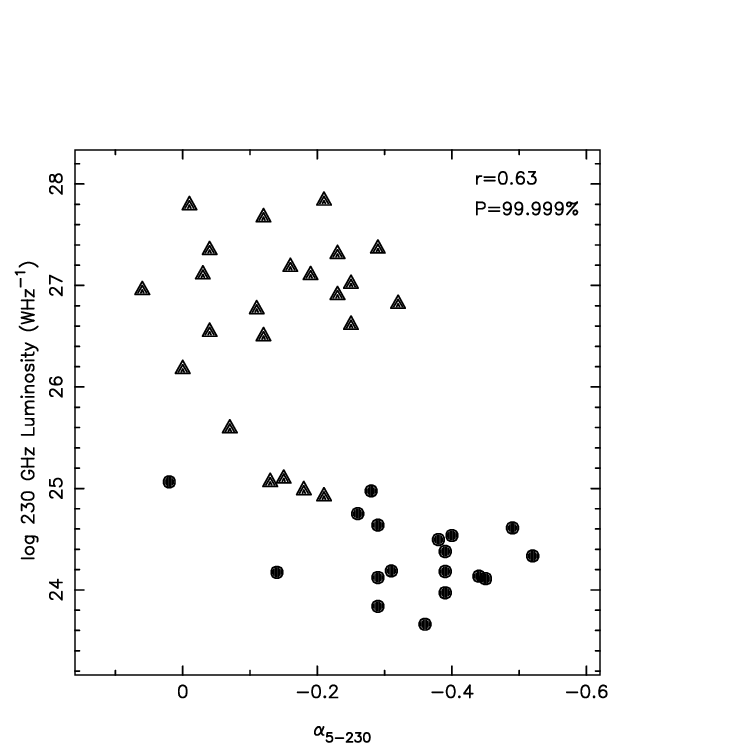}
\end{picture}
\caption[dum]{Correlation of $\alpha_{5-230}$ with 230~GHz
luminosity. Triangles are LBLs and circles are HBLs. The correlation
coefficient, $r$, and its significance, $P$, are indicated on the
plot.}
\end{figure}
\begin{figure}
\setlength{\unitlength}{1in}
\begin{picture}(3.5,2.5)
\includegraphics{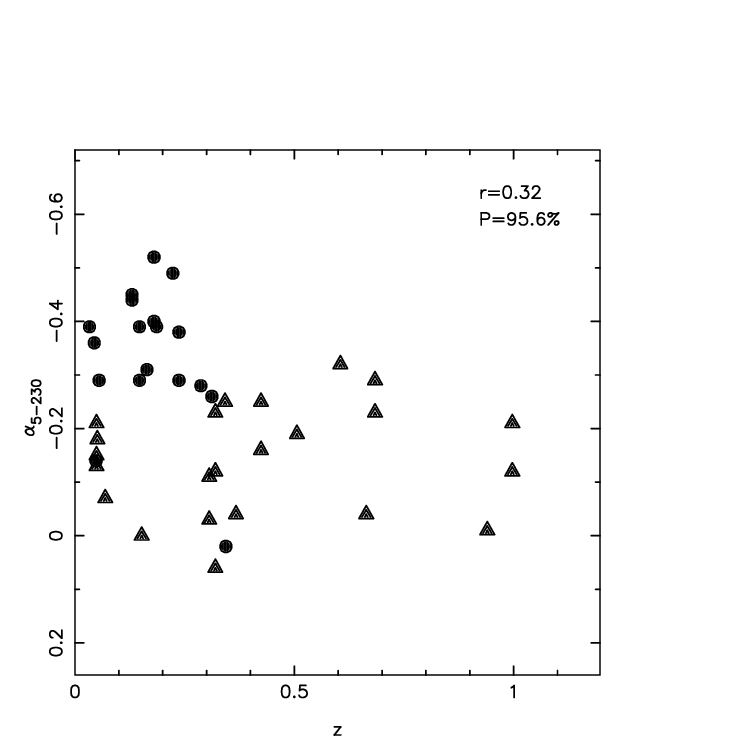}
\end{picture}
\caption[dum]{Correlation of $\alpha_{5-230}$ with redshift. 
Triangles are LBLs and circles are HBLs. The correlation
coefficient, $r$, and its significance, $P$, are indicated on the
plot.}
\end{figure}
\begin{figure}
\setlength{\unitlength}{1in}
\begin{picture}(3.5,2.5)
\includegraphics{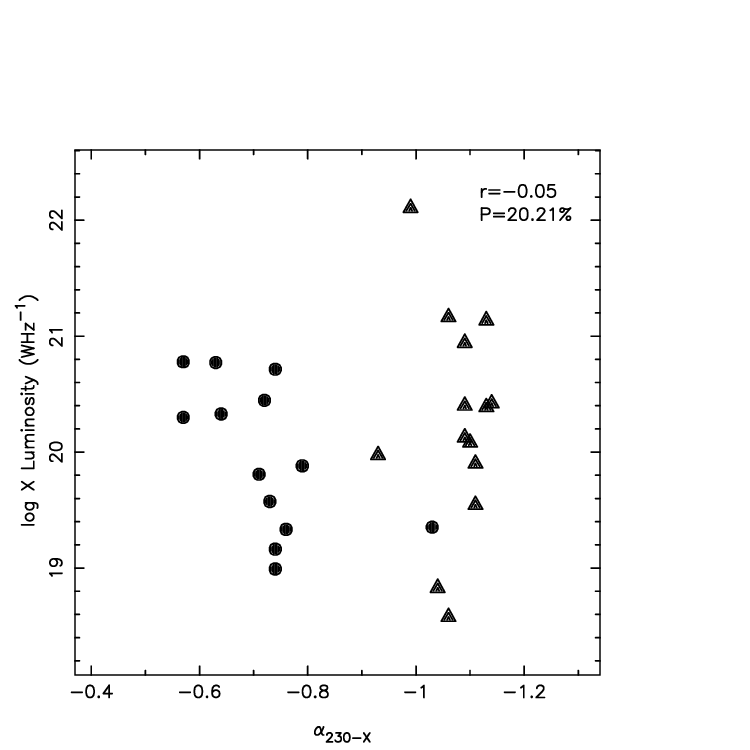}
\end{picture}
\caption[dum]{Correlation of $\alpha_{230-X}$ with X-ray
luminosity. Triangles are LBLs and circles are HBLs. The correlation
coefficient, $r$, and its significance, $P$, are indicated on the
plot.}
\end{figure}
\begin{figure}
\setlength{\unitlength}{1in}
\begin{picture}(3.5,2.5)
\includegraphics{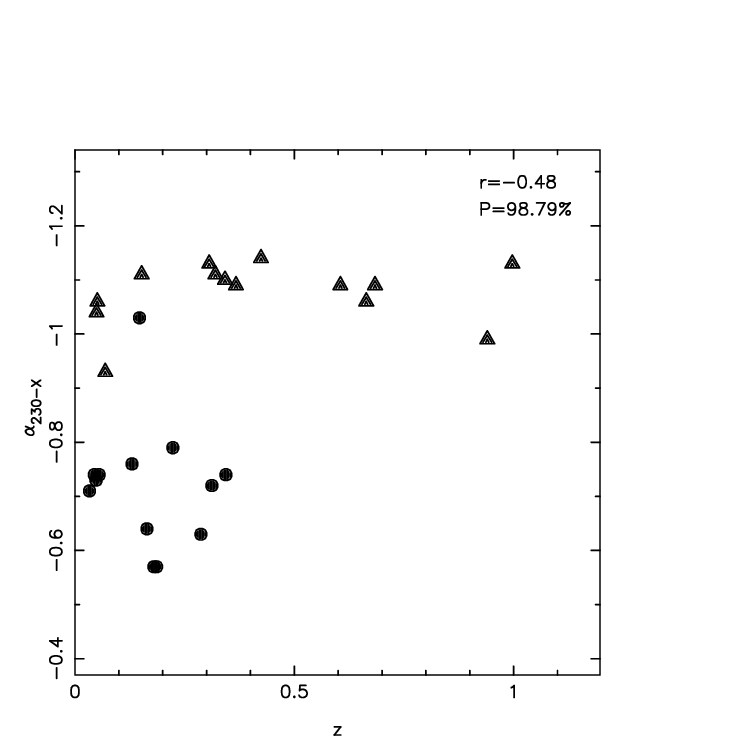}
\end{picture}
\caption[dum]{Correlation of $\alpha_{230-X}$ with redshift. 
Triangles are LBLs and circles are HBLs. The correlation
coefficient, $r$, and its significance, $P$, are indicated on the
plot.}
\end{figure}

Fig. 5 shows that $\alpha_{5-230}$ is indeed correlated with
230~GHz luminosity in the sense that more luminous objects have more
positive values of $\alpha_{5-230}$ (a Spearman rank-order correlation 
test gives $r=0.63$ at
the $P=>99.99$ per cent confidence level) whereas Fig. 6 shows no strong 
correlation between $\alpha_{5-230}$ and redshift ($r=0.32$, $P=95.6$ per
cent). Subsequently, the correlation shown in Fig.~5 does not
result from a correlation of luminosity with both redshift (not shown
but expected for flux limited samples) and spectral index. Possible
origins of this correlation are discussed in the next section.
It is thus found that {\em the radio--millimetre spectral indices of LBLs are
slightly steeper than those of HBLs}.

The $\alpha_{230-X}$ distributions (Fig. 4) are clearly different
($>$99.99 per cent confidence with a KS test). 
Mean values of
$\alpha_{230-X}$ are $-0.68\pm0.07$ for the HBLs and $-1.08\pm0.06$
for the LBLs. An important point to note is that the two intermediate sources 
that were discussed in Section~2 have intermediate
values of $\alpha_{230-X}$ (both fall in the $-0.9$ to $-1.0$ bin; not shown). 
This point is discussed further in the next section. 

Figs 7 and 8 show the correlations of $\alpha_{230-X}$ with
X-ray luminosity and redshift. Fig. 7 shows no correlation ($r=-0.05$,
$P=20.2$ per cent) whereas Fig. 8 shows a possible anti-correlation of spectral
index with redshift ($r=-0.48$, $P=98.85$ per cent) which it most likely an
artefact caused by the different redshift distribution of LBLs and HBLs 
(e.g. Gear et al. 1993b).  It is concluded that {\em the millimetre to X-ray
spectra of HBLs are significantly flatter than those of LBLs}. 

\section{Discussion}

The preceding analysis showed that there are probable differences
between the $\alpha_{5-230}$ distributions and distinct differences between the
$\alpha_{230-X}$ distributions of LBLs and HBLs. This section
discusses whether these results are consistent with a beaming model in which
the LBLs are aligned closer to the line-of-sight than the HBLs 
or whether intrinsic differences between the classes are required by the 
data. 

\subsection{The $\alpha_{5-230}$ distribution}

In Section 4 it was found that the HBLs have slightly steeper spectra
between 5 and 230~GHz than the LBLs and that this result might follow from the
correlation of $\alpha_{5-230}$ with 230~GHz luminosity. It is, however,
possible that this correlation is spurious and results from comparing two
groups of BL Lac objects with intrinsically different radio properties; 
note that
Fig. 5 shows little overlap between the two classifications. This section
investigates possible origins for the correlation assuming it is real.

One mechanism that can change both the luminosity and the spectrum is the
relativistic Doppler effect, i.e. the beaming model. 
These models were first devised (Browne 1983) to link LBLs with an unbeamed
parent population of Fanaroff-Riley class I radio galaxies (FRI; Fanaroff \&
Riley 1974). Subsequently, the HBLs, which display less extreme properties, 
were included in the same scheme but at larger viewing angles (less Doppler 
boosting) than the LBLs (Ghisellini \& Maraschi 1989). Various studies suggest
that the average angles to the
line-of-sight are $\sim10^{\rm{o}}$, 20$^{\rm{o}}$ and $60^{\rm{o}}$ for the 
LBLs, HBLs and FRIs
respectively (Urry, Padovani \& Stickel 1991; Ghisellini et al. 1993; Kollgaard
et al. 1996b).

The details of the correlation shown in Fig. 5 will
depend on the location of the synchrotron self-absorption turn-over frequency
with respect to the frequencies used to calculate the spectral index. More
specifically, a correlation will only be observed if the turn-over is Doppler
shifted through the observed frequency range.

The luminosity in the observer's frame, $L_{\nu}$, is related to that in the
source frame (denoted $'$) by $L_{\nu}=\delta^qL'_{\nu}$ where $\delta$ 
is the relativistic Doppler factor ($\delta =
[\Gamma_p-(\Gamma_p^2-1)^{1/2}\rm{cos}\,\theta]^{-1}$ where $\Gamma_p$ is the
pattern Lorentz factor and $\theta$ is the viewing angle). The exponent 
$q=3-\alpha$ for
a spheroidal source whereas $q=2-\alpha$ for a continuous jet. The former is
assumed below but note that the continuous jet case requires a higher Doppler
factor to produce a given luminosity. Using the same notation, the observed
frequency $\nu=\delta\nu'$.

The intrinsic synchrotron spectrum of a BL Lac object can be represented by
\begin{equation}
L'_{\nu'}=C'(\nu_1') \left ( \frac{\nu'}{\nu_1'} \right ) ^{\xi} \left \{
1-\rm{exp} \left[ - \left( \frac{\nu'}{\nu_1'} \right) ^{\alpha-\xi} \right]
\right\}
\end{equation}
where $\alpha$ is the optically thin spectral index, $\xi$ is the spectral
index in the optically thick regime, $\nu_1'$ is the frequency at which the
optical depth equals unity and $C'(\nu_1')$ is a constant. It follows that
\begin{equation}
L_{\nu}=\delta^{3-\alpha} C'(\nu_1') \left( \frac{\nu}{\delta\nu_1'} \right)
^{\xi} \left\{ 1-\rm{exp} \left[ - \left( \frac{\nu}{\delta\nu_1'} \right)
^{\alpha-\xi} \right] \right\}
\end{equation}

The intrinsic spectrum is constructed by assuming values for $\alpha$, $\xi$,
$\nu_1'$ and $C'(\nu_1')$. Equation (2) can then be used to calculate the 230
and 5 GHz luminosities, and hence $\alpha_{5-230}$, in the observer's frame 
as a function of $\delta$ (or $\theta$ if a value for $\Gamma_p$ is assumed). 
Fig.~9 shows several of these simple models plotted over the observed
correlation. The various parameters were chosen to give a good representation
of the data and good agreement with the unified scheme, these being
$\alpha=-0.6$, $\xi=0.4$ and $C'(\nu_1')=10^{25}$ WHz$^{-1}$; curves
corresponding to values of $\nu_1'$=3, 8, 12, 16 and 20 GHz are plotted from
right to left.
 Various values of
$\delta$ are shown on the solid curve ($\nu_1=8$~GHz). The corresponding values
of $\theta$ assume $\Gamma_p=5$ which is the approximate value indicated by 
both the measured superluminal speeds (Kollgaard et al. 1996b, and
references therein) and luminosity function studies (Urry \& Padovani 1995).

\begin{figure}
\setlength{\unitlength}{1in}
\begin{picture}(3.5,3.5)
\includegraphics{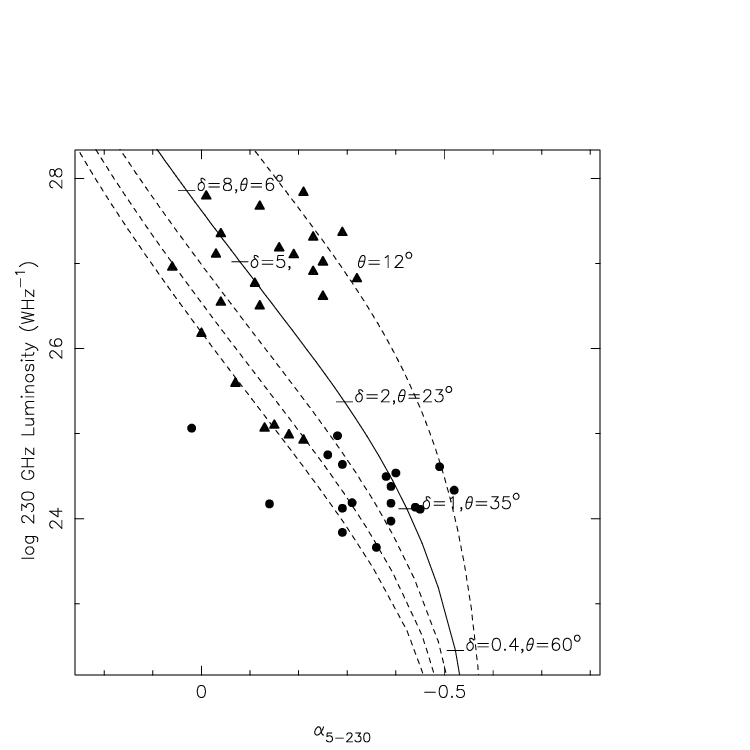}
\end{picture}
\caption[dum]{Simulated spectral index versus luminosity relationships plotted
over the data shown in Fig. 5. The lines represent values of
$\nu_1=3, 8, 12, 16, 20$ GHz (from right to left). 
Values of the relativistic Doppler factor, ($\delta$) and viewing angle
($\theta$ - see text) are shown on the solid curve for which $\nu_1=8$~GHz}
\end{figure}

The model clearly provides a good representation of the observed correlation
and is consistent with the unified scheme outlined above but are the assumed
parameters reasonable? The optically thin spectral index, $\alpha=-0.6$, is
close to that measured for large samples of radio sources (e.g. Conway,
Kellerman \& Long 1963) and thus
appears a valid assumption. As $\theta$
approaches 90$^{\rm{o}}$ the measured value of $\alpha_{5-230}$ will approach
$\alpha$ as indicated in Fig. 9 and so, for example, reducing $\alpha$ will
skew the envelope of models to the bottom right. Similarly increasing $\xi$
will skew the envelope to the top left.

The value of $\xi$ is difficult to constrain from existing
observations. For a simple homogeneous source $\xi=2.5$ but the composite 
spectra of BL Lac objects are
well known to be flatter than this in the radio regime. The
data for LBLs (Gear et al. 1994) give $\xi$ in the range 0.1--0.9 but
values towards both ends of this distribution give poor representations of the
observed correlation.

The value of $C'(\nu_1')$ fixes the various values of $\delta$
and $\theta$ along the curves. Observations of FRI radio galaxies at 230~GHz
would show whether the assumed value is reasonable but no data exist in the
literature at this time. However, estimates of $\delta$ for BL
Lac objects (mostly LBLs) exist in the literature from both Synchrotron
Self-Compton (SSC)
models (e.g. Ghisellini et al. 1993) and variability studies (Ter\"{a}sranta \&
Valtaoja 1994). The former study gives values of $\delta=0.01-14.3$
(mean=$3.9\pm4.2$; 33 objects) with only four objects having $\delta>10$ 
whilst the latter gives $\delta=0.2-4.3$ (mean=$1.4\pm1.2$; 11 objects). These
values are roughly consistent with those derived above and support the choice
of $C'(\nu_1')$.

It is thus concluded that the correlation of $\alpha_{5-230}$ with
230~GHz luminosity, if real, is consistent with that expected to arise if the
LBLs and HBLs are Doppler boosted FRI radio galaxies. The parameters of the
assumed model appear reasonable but are perhaps limited to a narrow range of
acceptable values.
However, it is important to note that such a simple model cannot be used to
infer that all of the differences between the spectra of BL Lac objects can be
explained within the framework of the beaming model. 
For example, Sambruna et al. (1996) showed that
whilst the accelerating jet model (Ghisellini \& Maraschi 1989) at a range of
viewing angles can reproduce successfully some of the spectral differences
between LBLs and HBLs, it does not lead to the observed wide range of peak
frequency or $\alpha_{ro}$.
These authors suggest that a difference in viewing angle coupled
with an intrinsic difference in jet properties, such as magnetic field strength
or jet size is required to explain the data. More recently, Fossati et
al. (1997,1998; 
see also Sambruna et al. 1996) have suggested that a link exists
between spectral shape and bolometric luminosity and this idea has been
extended to unify BL Lac objects in terms of viewing angle and jet electron
kinetic luminosity (Georganopoulos \& Marscher 1998). For the latter model, the
difference in luminosity between HBLs and LBLs is only partly attributed to a
difference in viewing angle. Furthermore, it can explain
the differences in redshift distribution and cosmological evolution between the
two classes but, as noted by the authors, does not include a treatment of
the inverse-Compton mechanism which may dominate the X-ray emission of 
LBLs (see next section).

\subsection{The $\alpha_{230-X}$ distribution}

The spectral index distributions between 230 GHz and the X-ray band are
significantly 
different and tend to cluster around values of $\sim-0.70$ for the HBLs and
$\sim-1.1$ for the LBLs although, as noted in Section~4, the two `intermediate'
BL Lac objects have intermediate values of $\alpha_{230-X}$ ($\sim-0.9$). More
millimetre observations of BL Lac objects that fall in the intermediate region
of the $\alpha_{ox}-\alpha_{ro}$ diagram are needed to confirm this trend.
The
$\alpha_{230-X}$ distributions show a more pronounced difference between the
two classes than the $\alpha_{rx}$ distributions. For example, Sambruna et
al. (1996) find $\alpha_{rx}=-0.57\pm0.06$ for the EMSS XBLs and
$\alpha_{rx}=-0.86\pm0.08$ for the `1 Jansky' RBLs with some overlap between
the two samples. This effect is likely to be
due to the influence of synchrotron self-absorption on the radio fluxes whereas
the 230 GHz emission is optically thin.

The clear dichotomy in $\alpha_{230-X}$ certainly cannot be
explained within the framework of the simple beaming model outlined in the
previous section but conversely cannot be used as evidence that beaming is
unimportant in these sources. 
As discussed above the beaming model in combination with
one or more intrinsic differences between the two classes may provide the best
fit to the data.

The data presented here as well as in the literature now strongly suggest 
that different mechanisms are responsible for the X-ray emission in the two
classes. 
Padovani \& Giommi (1996) have analysed the {\em ROSAT} X-ray spectra of a 
large sample of BL Lac objects. They find that the LBLs
are characterised by concave optical--X-ray continua and have flatter X-ray
spectra than the HBLs. The HBLs have convex overall broadband continua and the
X-ray spectral slopes are well correlated with $\alpha_{ox}$ and
anti-correlated with the X-ray to radio flux and cutoff frequency.
They conclude that the X-ray emission in HBLs is
produced by the synchrotron process whereas that from LBLs emanates from the
SSC mechanism. In this scenario the HBLs must have high-energy cutoffs to the
synchrotron emission at higher frequencies (UV--X-ray) than do the LBLs
(IR--optical) otherwise synchrotron emission would dominate in both cases.

The 230~GHz measurements presented here are on the high-frequency
side of the self-absorption turn-over and provide a measure of the optically
thin spectral index. The mean value of $-0.68\pm0.07$ for the
HBLs is strikingly similar to the canonical value for large samples of radio
sources (e.g. Conway et al. 1963) providing strong evidence that the X-rays are
produced by the synchrotron mechanism. 

For the LBLs there is no {\em a priori} reason to assume that the observed
spectral indices are not also produced by the synchrotron process but with a
different electron energy spectrum to the HBLs, although this situation appears
somewhat contrived. A more realistic interpretation, taking into account the
above discussion, is that the X-ray 
emission from the LBLs is inverse-Compton radiation. For the `intermediate'
sources it is natural to speculate that the X-ray emission comes from a
combination of both mechanisms, requiring that the X-ray synchrotron emission
is significantly steepened by radiative losses in these objects.

\section{Summary}

Millimetre wavelength observations of 22 XBLs of which 19 are HBLs, 1 is a LBL
and 2 are `intermediate' sources  
have been used along with similar published data for LBLs to search for
differences between the radio--millimetre--X-ray spectral energy distributions
of these objects. We find:

(1) The $\alpha_{5-230}$ two-point spectral indices of HBLs are slightly
    steeper than those of LBLs. Furthermore $\alpha_{5-230}$ is correlated with
    230~GHz luminosity. If this correlation is real, i.e. it is not a result of
    comparing two populations of BL Lac objects with intrinsically different
    radio properties, then it is consistent with existing unified
    schemes where BL Lac objects are FRI radio galaxies viewed along the jet
    axis, the HBLs being at larger viewing angles than the LBLs. The assumed
    parameters of this model, however, must take a relatively narrow range of
    values and it is unlikely that a difference in beaming parameters alone 
    can unify LBLs, HBLs and FRIs.

(2) The $\alpha_{230-X}$ two-point spectral indices of HBLs are significantly
    flatter than those of LBLs and have values consistent with those expected
    for synchrotron radiation. The X-ray emission from the LBLs is more likely
    to be produced by the inverse-Compton mechanism. Interestingly, the two
    `intermediate' BL Lac objects also have intermediate values of
    $\alpha_{230-X}$ suggesting that both processes contribute to the
    X-ray emission in these sources.

The analysis presented in this paper thus points to a unified model for BL Lac
objects that combines a spread of viewing angles with a spread of intrinsic
source parameters. In this respect the model of Georganopoulos \& Marscher
(1998), adapted to incorporate inverse-Compton losses appears promising.

\section*{ACKNOWLEDGMENTS}

The James
Clerk Maxwell Telescope is operated by the Joint Astronomy Centre in
Hilo, Hawaii on behalf of the parent organizations PPARC in the United
Kingdom, the National Research Council of Canada and The Netherlands
Organization for Scientific Research. This research has made use of
the NASA/IPAC Extragalactic Database (NED) which is operated by the
Jet Propulsion Laboratory, California Institute of Technology, under
contract with the National Aeronautics and Space
Administration. J.A.S. acknowledges the support of a PPARC PDRA.

\bsp
\end{document}